\documentclass[12pt,preprint]{aastex}
\begin{document}
 
\def\today{\number\year\space \ifcase\month\or  January\or February\or
        March\or April\or May\or June\or July\or August\or
September\or
        October\or November\or December\fi\space \number\day}
\def\fraction#1/#2{\leavevmode\kern.1em
 \raise.5ex\hbox{\the\scriptfont0 #1}\kern-.1em
 /\kern-.15em\lower.25ex\hbox{\the\scriptfont0 #2}}
\def\spose#1{\hbox to 0pt{#1\hss}}
\def\simlt{\mathrel{\spose{\lower 3pt\hbox{$\mathchar''218$}}
     \raise 2.0pt\hbox{$\mathchar''13C$}}}
\def\simgt{\mathrel{\spose{\lower 3pt\hbox{$\mathchar''218$}}
     \raise 2.0pt\hbox{$\mathchar''13E$}}}
\def\etal{et al. }
\def\simlt{\mathrel{\hbox{\rlap{\hbox{\lower4pt\hbox{$\sim$}}}\hbox{$<$}}}}
\def\simgt{\mathrel{\hbox{\rlap{\hbox{\lower4pt\hbox{$\sim$}}}\hbox{$>$}}}}
\def\kms{${\rm km\;s^{-1}}$}
\def\mps{${\rm m\;s^{-1}}$}

\title{\Large \bf Dispersing the Gaseous Protoplanetary Disc 
and Halting Type II Migration}

\author
{M. Lecar \& D. D. Sasselov}
\affil{Harvard-Smithsonian Center for Astrophysics, 60 Garden St.,
Cambridge, MA 02138}

\begin{abstract}
More than 30 extra-solar Jupiter-like planets have shorter
periods than the planet Mercury. It is generally accepted that they
formed further out, past the 'snow line' ($\sim$1~AU), and migrated
inwards. In order to be driven by tidal torques from the gaseous disc,
the disc exterior to the planet had to contain about a planetary mass.
The fact that the planets stopped migrating means that their
outer disc was removed. We suggest, following the simulation by
Bate \etal (2003), that the outer disc was accreted by the planet. This 
not only halts migration but removes the outer disc for planets interior
to about 2~AU. The disc further out could have been removed by
photoevaporation (Matsuyama \etal 2003). Furthermore, as also shown
by Bate \etal (op cit) this process also provides an upper limit to
planetary masses in agreement with the analysis of observed planetary
masses by Zucker \& Mazeh (2002).
In this scenario, the endgame is a race. The central star is accreting the
inner disc and the planet, while the planet is accreting the outer disc.
The planet survives if it accretes its outer disc before being accreted
by the star. The winner is determined solely by the ratio of the mass of
the outer disc to the local surface density of the disc. Some planets are
certainly eaten by the central star.
\end{abstract}
\keywords{extrasolar planets, Jupiter}

\section{Introduction}

Planets whose 'Roche Radii' are comparable to the disk scale height form a gap
in the disc. This was first suggested by Lin \& Papaloizou (1986) as the mechanism 
to limit the growth of Jupiter. It was thought that the gap acted like semi-
permeable membrane preventing gas from flowing inwards.
Because the close-in extra-solar planets had migrated inwards, and then 
stopped, they must have had an outer disc and then lost it. 
Previously, we suggested (Lecar \& Sasselov 1999) that gas from the outer disc would stream across the 
gap, when the gap width was less than the distance gas could travel in an 
orbital period. We thought that the gas would   
sneak across the gap when the planet was on the other side of the star, and 
join the inner disc. We speculated that 3-D disc simulations 
would confirm this. Bate \etal (2003) recently performed just 
the simulation we wished for. Gas did, indeed, stream across the gap, almost 
as if the gap wasn't there, but contrary to our speculation, did not reach the 
inner disc. Almost all of the gas was accreted onto the planet. 
However, this also halts migration. 
The accretion of the outer disc by the planet solves two problems:
stopping the migration and removing the outer disc. A recent study
of removal of the disc by photoevaporation showed that that process
is only effective exterior to 2.4~AU (Matsuyama \etal 2003).
In the minimum
mass solar nebula (Hayashi 1981) there are about four Jupiter masses of
gas interior to 2.4~AU. In any case, we have to account for the more than
50 extra-solar planets interior to 1 AU. Since the 'snow line' starts at
1 AU, there is a possibility that planets outside of 1 AU formed in place,
or did not migrate much.

An argument in favor of the planet accreting its outer disc is that it provides
a natural upper limit to the masses of planets of 10 Jupiter masses.
Zucker \& Mazeh (2002) found that to be the upper limit to planetary masses
and the lower limit to the masses of brown dwarfs. The formation of
gas giants by gravitational instability (Boss 2000) provides no natural
upper limit to the masses.

We now discuss in more detail the migration process, following Lin \&
Papaloizou (1986). We wish to illustrate when migration switches 
from Type~I (no gap) to Type~II (gap) (Ward 1997).
A planet with semi-major axis $a$ migrates at a rate
\begin{equation}
\dot{a}= -a{\Omega}{\mu}_{\rm pl}{\mu}_{\rm disk}I
\end{equation}
where
\begin{equation}
{\Omega}^2a^3= GM_{\odot};~~~{\mu}_{\rm pl}= \frac{M_{\rm pl}}{M_{\odot}};
~~~{\mu}_{\rm disk}=  \frac{2{\pi}{\Sigma}a^2}{M_{\odot}}.
\end{equation}
For a minimum mass solar nebula (Hayashi 1981), the surface mass density of the disk is
${\Sigma}=1700~g~cm^{-2}$ at 1 AU, and ${\Sigma}\propto x^{-3/2}$, where $x=a/AU$.
The dimensionless integral, $I$, is given by
\begin{equation}
\int_{{\Delta}_{out}} dx{\frac{(1+x)^{5/2}}{x^4}} -
\int_{{\Delta}_{in}} dx{\frac{(1-x)^{5/2}}{x^4}} \cong
\frac{1}{3}({\frac{1}{{\Delta}_{out}^3}-\frac{1}{{\Delta}_{in}^3}}) +
\frac{5}{2}({\frac{1}{{\Delta}_{out}^2}+\frac{1}{{\Delta}_{in}^2}})
\end{equation}
where for ${\Delta}\ll 1$, $I$ is insensitive to the upper limits.
If there is no gap, ${\Delta}_{in}={\Delta}_{out}$ and 
$I=\frac{5}{2}\frac{1}{{\Delta}^2}$, where ${\Delta}=\frac{h}{a}=\frac{c_s}{V}$,
$V$ is the circular orbital velocity, $c_s$ is the sound speed, and
$h$ is the scale height. Typically, ${c_s}/{V} \cong 0.06~x^{1/4}$.

If there is a gap, the planet moves within the gap to equalize
the torques from the inner and outer discs. This is accomplished
by
\begin{equation}
{\Delta}_{out}= {\Delta}({1+\frac{5}{4}{\Delta}}), ~~~~~~~~
{\Delta}_{in}= {\Delta}({1-\frac{5}{4}{\Delta}})
\end{equation}
Henceforth, the planet responds to the inward migration of the 
outer disc, which is driven by viscosity. Once a gap is opened,
the further evolution is controlled by viscous accretion and is referred to
as type~II migration (Ward 1997).

The mass accretion rate is
\begin{equation}
\dot{M}_{acc}= 2{\pi}{\Sigma}a\dot{a}\equiv 
2{\pi}{\Sigma}a^2{\frac{\dot{a}}{a}} \equiv
2{\pi}{\Sigma}a^2\frac{1}{t_{acc}},
\end{equation}
which by continuity is independent of $x$. If ${\Sigma}\propto x^{-3/2}$, then
$t_{acc}\propto x^{1/2}$. The time to accrete a Jupiter mass is 
\begin{equation}
t_{M_{\rm J}}= \frac{M_{\rm J}}{\dot{M}_{acc}}= 
\frac{M_{\rm J}}{2{\pi}{\Sigma}a^2}{t_{acc}}.
\end{equation}
If the outer disc has, say, a Jupiter mass, in order for a planet to accrete the
outer disc before being swept into the star, we require that
\begin{equation}
t_{M_{\rm J}}< t_{acc}
\end{equation}
or
\begin{equation}
\frac{M_{\rm J}}{2{\pi}{\Sigma}a^2}= 
\frac{M_{\rm J}}{2{\pi}{\Sigma}_0a_0^2x^{1/2}}< 1
\end{equation}
The migration halts when
\begin{equation}
x^{1/2}> \frac{M_{\rm J}}{2{\pi}{\Sigma}_0a_0^2}.
\end{equation}

For the minimum mass solar nebula, migration would halt
at $x=0.63$ or the orbital period $P=183$~days. For denser discs our
estimates (with ${\Sigma}_{min}=1700~g~cm^{-2}$ at 1 AU) are given in Table 1.
For surface densities larger than $4{\Sigma}_{min}$,
the planet is accreted by the star. 

More quantitatively, if the planet is at ${a}_0$, the ratio
$\frac{M_{\rm J}}{2{\pi}{\Sigma}_0a_0^2}$ can be written:
\begin{equation}
\int_{1}^{x} dx(x)^{1-n} = \frac{(x)^{2-n}}{2-n},
\end{equation}
if ${\Sigma}(x)= {\Sigma}(1)x^{-n}$ and $x=\frac{a}{a_0}$.
For the planet to accrete its outer disc before the star accretes the
planet, that quantity has to be less than 1.0, or $x\leq x_m$. This
is illustrated in Table 2.
For comparisom, if $a_0= 5.203~AU$ (Jupiter), Saturn is at $x=1.83$. If
$a_0= 1~AU$, photoevaporation is effective at $x> 2.4$.

So far, we have avoided a discussion of the physical source of the 
viscosity and the value of $\alpha$ (Shakura \& Sunyaev 1973), 
because our result is independent of 
this physics. But, to make contact with the literature, we note that
with our prescription (also used by Bate \etal, op cit), ${\alpha}\propto {\alpha}_0x^{1/2}$,
we require ${\alpha}_0\leq 3{\times}10^{-4}$ in order that the mass accretion be less
than $10^{-8}M_{\odot}{yr^{-1}}$. This is to limit the accretion luminosity
to yield 
the so called 'passive disc' (Chiang \& Goldreich 1997; Sasselov \& Lecar 2000).
With this value of $\alpha$, we have $t_{acc}\approx 10^5x^{1/2}$~years, 
which is short,
suggesting that ${\alpha}_0$ is smaller.

We conclude with some speculations about our Jupiter. In the minimum
mass solar nebula, with ${\Sigma}\propto x^{-3/2}$, there is about 2.7 Jupiter
masses of gas between Jupiter and Saturn. Clearly, Jupiter (and Saturn)
did not accrete it all. However the investigation of photoevaporation
of the outer disc was motivated by the fact that Saturn has only $\frac{1}{3}$rd
the mass of Jupiter, and in any case, $x^{-3/2}$ yields a divergent mass
($\propto x^{1/2}$).
The surface density profile must steepen. If we keep the minimum mass
surface density at Jupiter (143 g/cm$^2$), but allow the surface density
to decrease outwards at a steeper rate, say $x^{-7/2}$, then there is less 
than a Jupiter mass between Jupiter and Saturn. If we allow Jupiter to start 
accreting when its mass was $\geq 0.1~M_{\rm J}$, as suggested by Bate \etal 
(2003), then the outer disc is in Jupiter.

\bibliography{journals}

\begin{table}[!h]
\begin{center}
\begin{tabular}{|c|c|c|}
\hline
$\Sigma$    & x          & P \\
            & (a/AU)     & (days)  \\
\hline\hline
${\Sigma}_{min}$  & 0.630  & 183 \\
2${\Sigma}_{min}$ & 0.157  &  23 \\
3${\Sigma}_{min}$ & 0.070  &   7 \\
4${\Sigma}_{min}$ & 0.039  &   3 \\
\hline
\end{tabular}
\end{center}
\caption[]{Orbital distances $x$ and periods $P$ at which a planet
accreting its outer disc would stop migrating, as a function of disc
surface density $\Sigma$.}
\end{table}

\begin{table}[!h]
\begin{center}
\begin{tabular}{|c|c|c|}
\hline
$n$    &$\frac{(x)^{2-n}}{2-n}$   & $x_m$ \\
\hline\hline
3/2 & $2(x^{1/2}-1)$ & 2.25 \\
2   & $ln~x$         & 2.72 \\
5/2 & $2(1-x^{1/2})$ & 4.0  \\
3   & $(1-x^{-1})$   & $\infty$ \\
\hline
\end{tabular}
\end{center}
\caption[]{The minimum orbital distance, $x_m$, where migration halts,
for different types of discs.}
\end{table}

\end{document}